\documentclass[letterpaper]{article} 
\usepackage[submission]{aaai2026}  
\usepackage{times}  
\usepackage{helvet}  
\usepackage{courier}  
\usepackage[hyphens]{url}  
\usepackage{graphicx} 
\usepackage{natbib}  
\usepackage{caption} 
\usepackage[font=footnotesize,labelfont=bf]{caption}
\usepackage{booktabs}
\usepackage{makecell}

\usepackage{amsmath,amssymb,amsthm}
\usepackage{thmtools,cleveref}
\usepackage{thm-restate}
\usepackage{tikz}
\usetikzlibrary{arrows,positioning,shapes,decorations,automata,backgrounds,petri,fit,calc,patterns}

\usepackage{csquotes}

\usepackage{soul}
\usepackage{lipsum}
\usepackage{xcolor}
\usepackage{paralist}

\usepackage{mathtools}

\usepackage{enumitem}

\newlist{inlineenum}{enumerate*}{1}
\setlist*[inlineenum]{mode=unboxed,label=\alph*}

\usepackage{blindtext}
\usepackage{multicol}
\usepackage{pgf-pie}
\usepackage{pgfplots}

\tikzstyle{startstop} = [rectangle, rounded corners, minimum width=0.5cm, minimum height=0.5cm,text centered, draw=black, fill=white]
\tikzstyle{break} = [rectangle, minimum width=0.5cm, minimum height=0.5cm, text centered, draw=black, fill=white]
\tikzstyle{joinedprocess} = [rectangle, minimum width=0.5cm, minimum height=0.5cm, text centered, draw=black, fill=olive!30!green!10, thick]
\tikzstyle{soloprocess} = [rectangle, minimum width=0.5cm, minimum height=0.5cm, text centered, draw=blue!30!darkgray!90, fill=blue!45!white!15, double]
\tikzstyle{studentprocess} = [rectangle, minimum width=0.5cm, minimum height=0.5cm, text centered, draw=black, fill=yellow!20]
\tikzstyle{decision} = [diamond, aspect=6, minimum width=0.3cm, minimum height=0.4cm, text centered, draw=black, fill=white]
\tikzstyle{decisionsign} = [diamond, aspect=2, minimum width=0.01cm, minimum height=0.04cm, text centered, draw=black, fill=white]
\tikzstyle{arrow} = [thick,->,>=stealth]
\tikzstyle{line} = [thick,-,>=stealth]

\title{Thinking Like a Student: AI-Supported Reflective Planning\\
in a Theory-Intensive Computer Science Course}

\author {
   Noa Izsak
}
\affiliations{
   noa.izsak@cispa.de
}

\begin{document}

\maketitle

\begin{abstract}
In the aftermath of COVID-19, many universities implemented supplementary ``reinforcement" roles to support students in demanding courses. Although the name for such roles may differ between institutions, the underlying idea of providing structured supplementary support is common. However, these roles were often poorly defined, lacking structured materials, pedagogical oversight, and integration with the core teaching team.
This paper reports on the redesign of reinforcement sessions in a challenging undergraduate course on formal methods and computational models, using a large language model (LLM) as a reflective planning tool. 
The LLM was prompted to simulate the perspective of a second-year student, enabling the identification of conceptual bottlenecks, gaps in intuition, and likely reasoning breakdowns before classroom delivery. 
These insights informed a structured, repeatable session format combining targeted review, collaborative examples, independent student work, and guided walkthroughs. 
Conducted over a single semester, the intervention received positive student feedback, indicating increased confidence, reduced anxiety, and improved clarity, particularly in abstract topics such as the \emph{pumping lemma} and \emph{formal language expressive power comparisons}. 
The findings suggest that reflective, instructor-facing use of LLMs can enhance pedagogical design in theoretically dense domains and may be adaptable to other cognitively demanding computer science courses.

\smallskip
\noindent
{\textbf{Keywords:} AI in Education, Reflective Teaching, Formal Methods, Undergraduate CS, Conceptual Diagnosis.}
\end{abstract}

\section{Introduction}
\label{sec:introduction}

Following the COVID-19 pandemic, many universities introduced a new reinforcement roles, aimed to support students in academically demanding courses~\cite{jabeen2025transforming}. 
Although the name for such roles may differ between institutions, the underlying idea of providing structured additional support remains common. 
However, these roles, although well-intentioned, often were poorly defined, lacking structured materials, pedagogical oversight, and accountability. 
The reinforcement instructors were neither fully integrated into the core course staff nor sufficiently empowered to redesign the content or delivery methods. 
As a result, reinforcement sessions frequently failed to benefit the very students they were meant to serve.

This paper presents an experience report from a mandatory second-year undergraduate course on Formal Methods and Computational Models, a foundational yet abstract and challenging requirement for a bachelor's degree in computer science and software engineering majors.
The course is widely regarded by students as one of the most difficult in the curriculum, due to its abstract nature, rapid pace, and reliance on formal reasoning. 
Topics include deterministic and non-deterministic automata, regular and context-free languages, the pumping lemma, Turing machines, and decidability. 
Although this intervention did not alter the core syllabus or lecture content, it provided an independently structured reinforcement track targeting conceptual clarity and building confidence.

As a teaching assistant (TA) for several years, I had repeatedly witnessed student frustration with the existing reinforcement format. 
Weaker students struggled to keep up, while stronger students disengaged from what they perceived as repetitive and unhelpful sessions. 
When I was offered the opportunity to take on the role, I agreed on the condition that I could radically redesign the structure and goals of the sessions.

At the core of this redesign was a novel use of a large language model (LLM), specifically ChatGPT~\cite{kasneci2023chatgpt,wang2024large}, not as a tutor or answer generator, but as a \emph{reflective design assistant}.
The model was prompted to simulate the persona of a second-year student, and was updated weekly with the relevant lecture and tutorial material. 
By probing its responses with common questions, edge cases, and intentionally ambiguous phrasing, potential conceptual bottlenecks and areas where instructional clarity was lacking were identified, a process aligning with prior work on LLMs for conceptual diagnosis~\cite{kokver2025artificial,bewersdorff2023assessing}.
These insights shaped the structure, pace, and focus of each session, which was built around anticipated misunderstandings, intuitive gaps, and reasoning breakdowns revealed through this process.

Rather than relying on predefined slides or handouts, each session was constructed to directly target these anticipated difficulties.
The sessions combined formal reviews, collaborative examples, independent student attempts, and carefully timed walk-through.
The aim was to provide clarity and confidence to weaker students, while challenging stronger students through abstraction and critical questioning.

Although the intervention was conducted over a single semester with a modest sample size, the feedback from the participating students highlighted increased clarity, reduced anxiety, and a greater sense of mastery; particularly in areas traditionally viewed as stumbling blocks, such as the pumping lemma and expressive power comparisons.

This work aligns with the goals of the educational AI community by offering a concrete, instructor-led case of AI-supported teaching innovation in a non-AI subject area. 
The use of large language models here is not to teach AI, but to leverage it as a design-time assistant to improve instruction in theoretically dense domains. 
This reflective use of AI, aimed to emulating learner perspectives and thinking and guiding pedagogical decisions, represents a novel and underexplored approach that could be generalized to other cognitively demanding computer science courses.

The rest of the paper is organized as follows: Sec.\ref{sec:related} reviews related work on AI integration in computing education and the use of AI for conceptual diagnosis, which we situate within broader developments in computing curricula and educational AI practice. 
Sec.\ref{sec:design} describes the course context, the redesign of the reinforcement sessions, and the use of ChatGPT as a reflective planning tool. Sec.\ref{sec:evaluation} presents qualitative and quantitative findings from student feedback. Then Sec.\ref{sec:discussion} discusses the broader implications and potential for adaptation to other courses, lastly, Sec.\ref{sec:conclusion} concludes the paper.

\section{Related Work}
\label{sec:related}

\paragraph{AI in the Computer Science Curriculum.}
The \emph{CS2023-curriculum} \cite{kumar2024computer, eaton2024artificial} emphasizes the integration of AI-related reasoning skills and applications into undergraduate computing programs. 
This reflects a broader trend where AI tools are no longer confined to specialized courses, but rather integrated into foundational computing topics. 
Though examples of AI integration exist in advanced courses such as autonomous agents \cite{rosenthal2023autonomous}, comparable efforts in theory-heavy subjects remain rare.

\paragraph{LLMs in Computing Education.}
Recent studies have examined the role of LLMs in supporting programming education and computing skills. 
Various works have shown their utility in improving error messages for students in the introductory stage of programming \cite{leinonen2023using}, providing automated feedback on experimentation \cite{bewersdorff2023assessing}, and supporting AI-assisted teaching practices \cite{wang2023exploring}.
Comprehensive surveys \cite{wang2024large,gan2023large} and critical reviews \cite{kasneci2023chatgpt,giannakos2025promise} identify the potential of generative AI, but also stress the importance of pedagogically grounded integration.

\paragraph{Misconception and Conceptual Detection.}
A growing body of work uses artificial intelligence to identify and address conceptual gaps in student learning. 
Natural language processing has been applied to detect misconceptions in educational contexts \cite{kokver2025artificial}, while explainable AI models have been proposed to identify logical errors in code \cite{hoq2025automated}. 
Meta-analyses show that AI can improve programming outcomes \cite{alanazi2025influence}, though its direct impact on conceptual understanding is more variable. 
These studies focus largely on reactive support; while our approach adapts these capabilities to the use of proactive design stages during the planning of instructional activities.

\paragraph{Gap and Contribution.}
Although prior work has demonstrated the value of LLM and AI tools in computing education, the application of such tools to instructional design in theoretical CS courses remains underexplored. 
This work addresses that gap by positioning the LLM not as a tutor or grader, but as a reflective partner for instructors, aimed to anticipate students' challenges and design structured and conceptually tailored reinforcement sessions.

\section{Design of the Intervention}
\label{sec:design}

\subsection{Course Context}
The intervention took place in a mandatory second-year undergraduate course on Formal Methods and Computational Models, covering topics such as: deterministic and non-deterministic automata, regular-, context-free-, and context-sensitive languages, the pumping lemma, Turing machines, and selected complexity topics. 
The course is widely regarded by students as one of the most conceptually challenging courses in the curriculum, due to its abstract formalisms, emphasis on proofs, and fast-paced progression.

In addition to weekly lectures and tutorials, the department offered supplementary ``reinforcement" sessions intended to provide additional practice and clarification. 
As noted in Sec.\ref{sec:introduction}, though the terminology of such roles varies across institutions, their shared purpose is to support students mastering difficult material through structured supplementary study hours.

\subsection{Pre-Redesign Challenges}
Prior to the redesign, reinforcement sessions were loosely structured, with content chosen ad hoc based on perceived difficulties or student questions from the preceding week. 
This reactive approach often resulted in:
\begin{itemize}
    \item \textbf{Fragmented coverage:} Core concepts were reinforced inconsistently, leaving gaps in cumulative understanding.
    \item \textbf{Variable depth:} Some topics received superficial review, while others were explored in disproportionate detail depending on immediate student concerns.
    \item \textbf{Limited sustained engagement:} Attendance tended to peak only before assessments, reducing opportunities for continuous reinforcement and practice.
\end{itemize}
Student feedback indicated that while such sessions could help with specific problems, they lacked a predictable structure to build long-term conceptual mastery.

\subsection{Redesign Principles}
The redesigned reinforcement sessions were guided by three integrated principles:
\begin{enumerate}[label=\arabic*.]
    \item \textbf{Anticipate and address misconceptions:} before they cause cumulative confusion, drawing on known points of difficulty in theoretical computer science.
    \item \textbf{Balance accessibility and challenge:} by providing a solid conceptual baseline for struggling students while engaging stronger students in higher-order reasoning.
    \item \textbf{Establish structure and predictability:} through a clear, repeatable session format that students could prepare for in advance, while modernizing examples and analogies in line with the \emph{CS2023} recommendations~\cite{kumar2024computer,eaton2024artificial}.
\end{enumerate}

\subsection{Use of ChatGPT as a Reflective Planning Tool}
A central innovation in the redesign was the use of ChatGPT as a \emph{reflective planning partner} rather than as a direct instructional tool for students. 
This framing aligns with calls to employ large language models (LLMs) not only as student-facing tutors, but as instructor-facing design aids.

Before each session, the model was updated with relevant lecture and tutorial content and instructed to simulate the perspective of a second-year student at the institution. 
By posing common, tricky, or edge-case questions, the LLM has been used to:
\begin{itemize}
    \item Identify conceptual dependencies that were not explicitly explained in the official slides or lecture notes.
    \item Detect subtle ambiguities in notation or terminology that might cause confusion.
    \item Suggest alternative explanations, analogies, or metaphors to support intuition building for abstract topics.
    \item Propose variations of problems for use in guided and independent practice.
\end{itemize}

Note that these insights were never shown directly to students but were embedded in instructional decisions. 
For example, when the LLM repeatedly conflated different closure properties, signaled the need to explicitly anchor terminology during review.
Similarly, incorrect applications of the pumping lemma to specific context-free grammar scenarios allowed the instructor to preemptively address common misconceptions.

Importantly, all AI-generated suggestions were critically reviewed, adapted, and integrated by the instructor to ensure alignment with course objectives and conceptual correctness, consistent with the pedagogical caution recommended in prior studies \cite{giannakos2025promise, kasneci2023chatgpt, wang2023exploring}. 
No student data or identifiable information was shared with the AI tool.
This process was informed by the recognition that large language models can produce inaccurate or biased responses, and that human oversight is essential for maintaining both factual accuracy and pedagogical integrity.

\begin{figure}[ht]
    \centering
    \scalebox{0.915}{\begin{tikzpicture}[node distance=1.0cm]

\node (start) [startstop] {\footnotesize Start of Session};
\node (present) [soloprocess, below of=start] { \footnotesize Review of prerequisites};
\node (easy) [joinedprocess, below of=present] { \footnotesize Solve and Discuss Collectively};

\node (present2) [soloprocess, below of=easy] { \footnotesize Present a Scaffolded Exercise};
\node (indep) [studentprocess, below of=present2] { \footnotesize Students Tackle an Exercise Independently};
\node (vol) [joinedprocess, below of =indep] {\footnotesize Volunteers Share \& Discuss Solutions};

\node (intuition) [decision, below of=vol] { \scriptsize {Understanding/Intuition?}};
\node (rintuition) [right of=intuition, right=1.5cm] {};
\node (difficulties) [decision, below of=intuition, below =-0.15] { \scriptsize {Significant Difficulties?}};
\node (presentSol) [soloprocess, below of=difficulties] { \footnotesize Present a Pre-constructed Solution Walkthrough};
\node (more) [decision, below of=presentSol] { \scriptsize More Exercises?};
\node (summary) [soloprocess, below of=more] { \footnotesize Summarize key points};
\node (end) [startstop, below of=summary] { \footnotesize End of Session};

\draw [arrow] (start) -- (present);
\draw [arrow] (present) -- (easy);
\draw [arrow] (easy) -- (present2);
\draw [arrow] (present2) -- (indep);
\draw [arrow] (indep) -- (vol);
\draw [arrow] (vol) -- (intuition); 
\draw [arrow] (intuition.west) --node[yshift=0.15cm]{\tiny \textbf{Yes}} ++(-0.95,0)  -- ++(0,-3.3) -- (more.west);    


\draw [arrow] (intuition) --node[xshift=-0.3cm]{\tiny \textbf{No}} (difficulties);
\draw [arrow] (difficulties.west) --node[yshift=0.15cm]{\tiny \textbf{No}} ++(-0.65,0)  -- ++(0,-2.0) -- (more.west);
\draw [arrow] (difficulties) --node[xshift=-0.3cm]{\tiny \textbf{Yes}} (presentSol);
\draw [arrow] (presentSol) -- (more);
\draw[arrow] 
  (more.east) --node[yshift=0.15cm]{\tiny \textbf{Yes}} ++(1.35,0)   -- ++(0, 6.3)   -- (present2.east); 
\draw[arrow] 
  (more.east) -- ++(1.35,0)   node[rotate =90, yshift=0.15cm, xshift=3cm]{\scriptsize \emph{next in progression}}-- ++(0, 6.3)   -- (present2.east); 
\draw [arrow] (more) --node[xshift=-0.3cm]{\tiny \textbf{No}} (summary);
\draw [arrow] (summary) -- (end);


\node[draw=black, thick, double, above=1.75cm of start, anchor=north] (legendbox) {
  \begin{tikzpicture}[baseline]

    \node (dec) [decisionsign] { ~ };
    \node[below of= dec, below=-0.75cm] {\small Decision};
    \node (s) [startstop, right of=dec, right = 0.25cm] {~};
    \node[below of= s, below=-0.75cm] {\small Start / End};
    \node (inst) [soloprocess, right of=s, right = 0.25cm] {~ ~};
    \node[below of= inst, below=-0.75cm] { \small Instructor};
    \node (stud) [studentprocess, right of=inst, right = 0.25cm] {~ ~};
    \node[below of= stud, below=-0.75cm] { \small Students};
    \node (join) [joinedprocess,right of= stud, right = 0.25cm] {~ ~};
    \node[below of= join, below=-0.75cm] { \small \shortstack[c]{Instr \& Stud}};
\end{tikzpicture}
};
\end{tikzpicture}}

    \caption{Flowchart diagram of session structure, indicating via color coding at each step whether the activity was led jointly by students and instructor, exclusively by students, or exclusively by the instructor.}
    \label{fig:flowchart}
\end{figure}
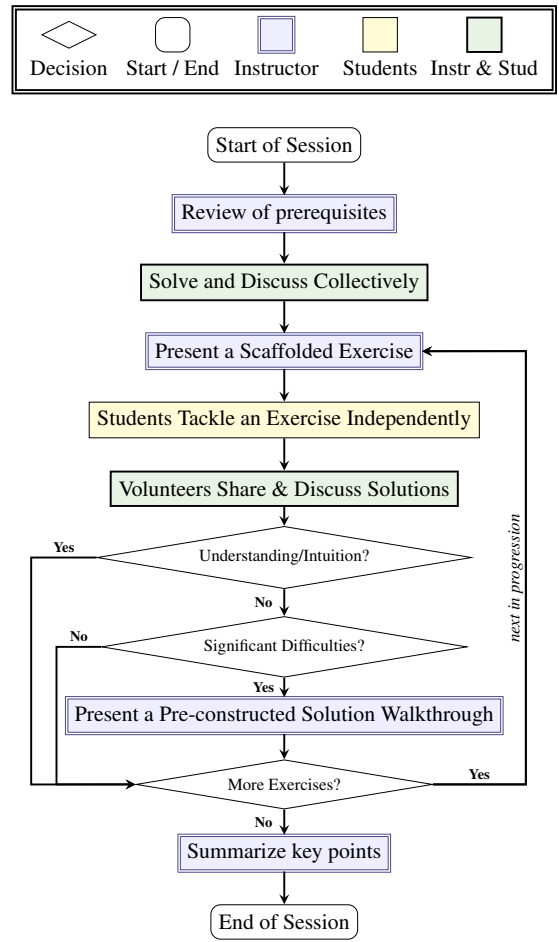

\subsection{Session Format}
Each session followed a repeatable, structured format (illustrated at Fig.\ref{fig:flowchart}) designed to support both foundational reinforce core concepts while cultivating higher-order reasoning.
The format combined instructor exposition, collaborative discussion, and independent student work --- each serving a deliberate pedagogical role.

\paragraph{First Academic Hour:}
\begin{enumerate}[label=\roman*.]
    \item Structured review of prerequisite knowledge to establish a common baseline for active participation.
    \item Instructor-guided examples or theorems, solved collectively with explicit attention to building intuition, surfacing common misunderstandings, and modeling reasoning strategies.
\end{enumerate}

\paragraph{Second Academic Hour:}
\begin{enumerate}[label=\roman*.]
    \item Presentation of progressively scaffolded exercises targeting the session's core topic.
    \item Brief independent work periods to promote active engagement and self-assessment.
    \item Volunteer-led sharing of solutions, with the instructor expanding, clarifying, or visualizing responses.
    \item Delivery of pre-constructed walkthroughs (often with diagrams or animations) when significant confusion persisted, targeting issues identified through prior sessions or through the LLM analysis.
\end{enumerate}
Session materials were shared in advance \emph{without solutions}, allowing students to preview upcoming content while retaining incentive to attend and participate, as well as to reflect on the problems before class.
Annotated solutions were provided afterward to reinforce learning. 
Visual aids, including hand-drawn figures and live illustrations, were used extensively, particularly for abstract topics like the pumping lemma and Turing machine encodings.
Printed handouts were used early in the semester for accessibility, later replaced with digital resources shared in real time.

\begin{figure}[t]
\centering
\begin{tikzpicture}
\begin{axis}[
    ybar, 
    bar width=14pt,
    width=0.995\linewidth,
    height=7cm,
    ymin=0, ymax=5.1, ytick={0,1,2,3,4,5},
    enlarge x limits=0.25,
    ylabel={Average Confidence (1--5)},
    symbolic x coords={Cardinality, REG, CFG, TM},
    xtick=data,
    xticklabel style={font=\normalsize, rotate=0, anchor=center, yshift=-6},
    legend style={at={(0.5,1.05)},anchor=south,legend columns=-1, font=\normalsize},
    ymajorgrids=true,
    grid style=dashed,
    ticklabel style={font=\normalsize},
    label style={yshift=-10, font=\normalsize},
    nodes near coords,
    nodes near coords align={vertical},
    nodes near coords style={fill=white!30,
    inner sep=0.1pt, 
    font=\bfseries\small,
    rounded corners=1pt,
    yshift=0.1cm
    },
    point meta=y
]

\addplot+[
    ybar,
    pattern=north east lines,
    pattern color=black,
    draw=black,
    error bars/.cd,
    y dir=both,
    y explicit
] coordinates {
    (Cardinality, 2.211) +- (0, 1.196)
    (REG, 3.111) +- (0, 0.809)
    (CFG, 3.188) +- (0, 0.950)
    (TM, 2.778) +- (0, 1.133)
};

\addplot+[
    ybar,
    fill=black!70,
    draw=black,
    error bars/.cd,
    y dir=both,
    y explicit
] coordinates {
    (Cardinality, 3.895) +- (0, 0.718)
    (REG, 4.222) +- (0, 0.786)
    (CFG, 3.875) +- (0, 0.781)
    (TM, 3.444) +- (0, 1.165)
};

\legend{Before Sessions, After Sessions}
\end{axis}
\end{tikzpicture}
\caption{Average self-reported confidence before and after AI-informed reinforcement sessions, with standard deviation error bars ($N=24$ maximum). Gains were observed across all topics, particularly in Cardinality and Turing Machines.}
\label{fig:confidence-chart}
\vspace{-1mm}
\end{figure}
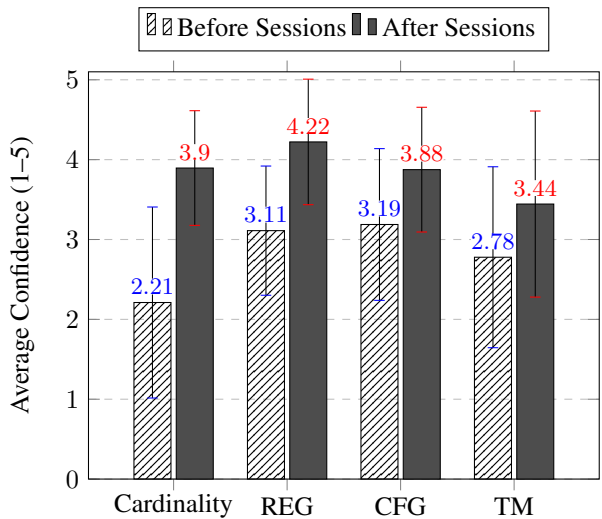

\section{Evaluation}
\label{sec:evaluation}

The redesigned reinforcement sessions were evaluated through an \emph{anonymous} survey, accessible to students throughout the semester to allow updates to their self-reported before/after confidence levels and related feedback.
The results reported here reflect the final responses collected at the end of the semester.
The goal was to capture both quantitative trends in self-reported confidence and qualitative perceptions of the sessions' value. Twenty-four students completed the survey, representing the majority of those regularly attending the sessions. While the sample size limits statistical generalization, the results provide meaningful directional insight into the intervention's impact.

This reflective structure aimed to support both weaker students; by providing clarity and confidence through guided review, as well as stronger students; by encouraging abstraction and critical questioning. 
Notably, a student who had shown little engagement early in the semester began posing increasingly abstract questions by midterm, including a complex inquiry about the relationship between the context-free grammar pumping lemma and single-variable grammars. This growth in reasoning was facilitated, in part, by the environment of structured challenge and collaborative problem-solving.

\begin{figure}[t]
    \centering
        \scalebox{0.6125}{\begin{tikzpicture}
    \pie[/tikz/nodes={text=white, font=\bfseries\LARGE},
/tikz/every pin/.style={align=center, text=black, font=\LARGE},
text=pin,
rotate=180,
explode=0.175,
color={green!70!cyan!20!olive!70, olive!50!blue!40!green!40,cyan!60!blue!50,red!70!blue!50, red!60!white!70,orange!50}
]{16/Essential, 26/Helpful, 37/Somewhat, 21/Minimal, 0/None}
    \end{tikzpicture}}
   
     \begin{center}
       \textbf{\small (a) Perceived Session Helpfulness}
    \end{center} 

\vspace{2mm}

     \centering
    \scalebox{0.6125}{\begin{tikzpicture}
    \pie[/tikz/nodes={text=white, font=\bfseries\LARGE},
/tikz/every pin/.style={align=center, text=black, font=\LARGE},
text=pin,
rotate=90,
explode=0.15,
color={olive!40!green!50,cyan!60!blue!50,red!60!white!70,orange!50}
]{27/High, 63/Moderate, 10/Low}
    \end{tikzpicture}}
    \begin{center}
         \textbf{\small (b) Session Impact on Confidence}
    \end{center}
    \caption{}\label{fig:session-help-a}\label{fig:session-confidence-b}
\end{figure}

\subsection{Quantitative Trends}
To measure perceived conceptual gains, students rated their confidence \emph{before} and \emph{after} the sessions on four key topics: Cardinality, Regular Languages and Regular Expressions (REG), Context-Free Grammars (CFG), and Turing Machines (TM). 
The averages and standard deviations scores (from 1 to 5) are shown in Figure~\ref{fig:confidence-chart}. 
Across all topics, mean confidence increased, with the most substantial relative gains observed in Cardinality and Turing Machines -- topics that students had rated lowest before the sessions. 
The standard deviations narrowed after the sessions for most topics, indicating greater alignment in perceived understanding.

\subsection{Categorical Perceptions}
\begin{figure}[t]
    \centering
    \centering
   \scalebox{0.6125}{\begin{tikzpicture}
    \pie[/tikz/nodes={text=white, font=\bfseries\LARGE},
/tikz/every pin/.style={align=center, text=black, font=\LARGE},
text=pin,
rotate=220,
explode=0.15,
color={olive!40!green!50,cyan!60!blue!50,red!60!white!70,orange!50}
]{60/Definitely, 30/Possibly, 10/No}
    \end{tikzpicture}}

    \caption{\textbf{\small Would You Recommend These Sessions?}}
    \label{fig:session-recommendation}
\end{figure}
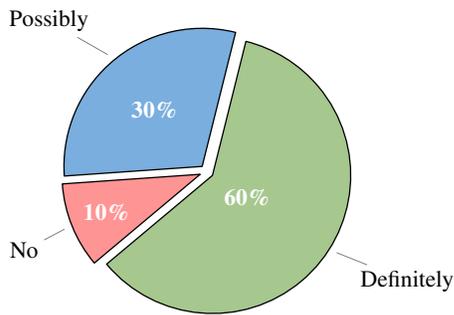

Three additional categorical survey questions provided a broader view of perceived value:
\begin{itemize}
    \item \textbf{Perceived Session Helpfulness}, \Cref{fig:session-help-a}a. 
    \item \textbf{Session Impact on Confidence}, \Cref{fig:session-confidence-b}b.
    \item \textbf{Recommendation Likelihood}, \Cref{fig:session-recommendation}
\end{itemize}
These categorical results align with the quantitative trends, suggesting that most students found the sessions moderately to highly beneficial, with a notable proportion perceiving them as essential to their learning.

\subsection{Qualitative Insights}
Throughout, students expressed appreciation for the slow pacing, the opportunity to ask clarifying questions, and the use of visual aids. The deliberate separation of instructional phases to: instructor-led, student-only, and joint, created a rhythm that helped manage cognitive load.

\noindent
Thematic analysis of open-ended responses identified four recurring themes:
\begin{itemize}
    \item \textbf{Reinforcement of foundational knowledge:} Students valued the chance to revisit prerequisite material, reporting that it eased anxiety when tackling advanced topics.
    \item \textbf{Clear structure and pacing:} The deliberate progression from review, to guided examples, to independent work was viewed as essential for deepening understanding.
    \item \textbf{Impact of visual explanations:} Many credited diagrams, animations, and live illustrations with making abstract concepts (such as the pumping lemma and Turing machine encoding) more tangible.
    \item \textbf{Supportive learning environment:} Students emphasized that the sessions encouraged open questioning without fear of judgment, fostering active engagement.
\end{itemize}

\paragraph{Representative student comments included:}

\begin{displayquote}
\enquote{The session organized the material clearly.}
\end{displayquote}

\begin{displayquote}
\enquote{There were questions that really helped solidify the basic concepts.}
\end{displayquote}

\begin{displayquote}
\enquote{The instructor answered every question and was clearly committed to improving the sessions.}
\end{displayquote}

\begin{displayquote}
\enquote{Explanations were clear and highlighted places where many students get confused.}
\end{displayquote}

\subsection{Limitations}
Two limitations are worth noting. First, the small sample size (ranging from 9 to 24 responses per topic) constrains statistical generalization. 
Second, as participation in the sessions was voluntary, the results may reflect a self-selection bias toward more motivated or engaged students. 
Nevertheless, the convergence of quantitative improvement, categorical endorsement, and qualitative feedback suggests that the redesigned reinforcement sessions had a meaningful positive impact on student learning experiences.

\section{Discussion}
\label{sec:discussion}
The findings from this intervention suggest that structured, AI-informed reinforcement sessions can play a valuable role in theory-intensive computer science courses.
While the original motivation was to address persistent challenges in a specific offering of a computational models course, the underlying design principles: structured scaffolding, deliberate pacing, targeted conceptual clarification, and reflective use of AI tools --- are not domain-specific. 
They may therefore be adaptable to a variety of contexts where abstract reasoning and formalism are central.

\looseness=-1
\subsection{Implications for Instructional Design}
Two aspects of the redesign appear particularly impactful. First, separating the sessions into distinct phases: {instructor-led review}, {collaborative exploration}, and {individual problem-solving}, helped manage cognitive load and sustain engagement. 
Second, using a large language model (LLM) as a \emph{reflective planning partner}--rather than a direct tutor--proved notably effective: several questions surfaced by the LLM aligned precisely with the actual difficulties reported by many students, and some explanations it inspired elicited the spontaneous \emph{``ahhh, now I get it"} reaction during sessions.
This alignment between AI-generated prompts and real student needs helped the instructor anticipate challenges, select more resonant examples, and refine pacing. 

This approach is especially relevant in domains where concepts are precise and cumulative, and where early misunderstandings can cascade into persistent knowledge gaps.
Recognizing that AI tools can produce inaccurate or misleading suggestions, all questions, examples, and pedagogical suggestions generated by the AI were \emph{carefully reviewed and validated} prior to integration.
This process ensured both conceptual correctness and alignment with the intended learning objectives.

The observed improvement in post-session confidence, combined with student emphasis on clarity, visual aids, and inclusivity, suggests that the combination of structured pedagogy and AI-informed preparation can enhance learning in theory-intensive courses.
Importantly, the AI component supported, rather than replaced, the instructor's expertise, maintaining academic rigor while introducing modern relevance and increasing accessibility.

\subsection{Potential for Adaptation}
\looseness=-2
Although the institutional term ``reinforcement" may vary, the concept of structured supplementary instruction can be applied across theory-heavy subjects such as algorithms, discrete mathematics, as well as programming-intensive courses.
Adaptation requires preserving the core scaffolding and misconception-anticipation strategies while tailoring examples and pacing to the specific domain.
Scaling such an approach may require additional instructor preparation time and access to AI tools that can be customized to the course context. 
However, once refined prompts and materials are established, much of the structure can be reused or adapted across offerings. 
Incorporating peer- or near-peer facilitation, supported by AI-informed preparation, could further expand reach without sacrificing instructional quality.

\subsection{Broader Educational Context}
This work illustrates how AI can serve as a catalyst for instructional reflection and design renewal, complementing rather than replacing instructor expertise.
In this framing, AI supports professional development by helping educators anticipate conceptual bottlenecks and modernize examples during the planning stage --- a role that is both distinct from and complementary to the more common student-facing applications of large language models.

\begin{table}[t]
\caption{Comparison between common \emph{student-facing} uses of large language models with the \emph{instructor-facing, design-time} approach used in this work, which shifts AI from a real-time answer generator to a reflective planning partner for more targeted scaffolding, pacing, and conceptual clarity in theory-intensive courses.}
\label{tab:ai_roles}
\centering
\renewcommand{\arraystretch}{1.65}
\begin{tabular}{|p{0.15\linewidth} p{0.345\linewidth} p{0.345\linewidth}|}
\hline
 & \textbf{Student-Facing} & \makecell[c]{\textbf{Instructor-Facing}\\ { \small \textbf{(This Work)}}} \\[2mm] \hline
\textbf{Role} 
& Tutor, Q\&A 
&  \makecell[l]{Planning aid,\\ diagnostic tool} \\[2mm] \hline
\textbf{Risks} 
& \makecell[l]{Overreliance \\ Unverified answers}
& \makecell[l]{Misalignment if \\ outputs not reviewed} \\[2mm]\hline
\textbf{Benefits} 
& Quick help
& \makecell[l]{Anticipates gaps \\ Refines pacing} \\
\hline
\end{tabular}
\end{table}

As summarized in Table~\ref{tab:ai_roles}, the reflective, instructor-facing use of AI adopted here differs substantially from conventional tutoring or automated Q\&A systems. 
While student-facing tools focus on immediate, reactive feedback during learning activities, the approach in this work positions AI as a design-time planning aid: outputs are used to predict likely misunderstandings, refine scaffolding, and select resonant examples, with all content critically reviewed before integration into instruction. 
This shift in focus highlights AI's potential as a behind-the-scenes partner in teaching innovation, enabling scalable improvements in clarity, engagement, and pacing without altering core course objectives.

The potential impact of instructor-facing AI extends beyond individual classrooms to curriculum design, faculty training, and cross-disciplinary applications in STEM subjects where precision and cumulative reasoning are critical.

\subsection{Practical Guidelines for Instructors}
For instructors in theory-heavy or abstract courses, three practical recommendations emerge:

\begin{enumerate}[label=\roman*.]
    \item \textbf{Embed AI at the design stage} Use a LLM to simulate a learner (i.e., a ``novice") reasoning, surface potential misconceptions, and test edge cases before teaching.
    \item \textbf{Maintain critical oversight} Treat AI output as a hypothesis to be verified, not a source of authoritative content.
    \item \textbf{Preserve structured scaffolding} Combine AI-informed insights with a predictable instructional sequence that includes review, guided exploration, and self practice.
\end{enumerate}
These guidelines balance innovation with caution, offering a pathway for instructors to adopt AI in a controlled, pedagogically aligned manner.

\section{Conclusion}
\label{sec:conclusion}
This work highlights how large language models can be used not only to generate content or support student learning, but also to enrich instructor preparation.
In the context of a mandatory second-year course on Formal Methods and Computational Models (a domain known for its abstract nature and steep learning curve) the integration of AI-informed planning with a structured pedagogical design led to observable improvements in student confidence, engagement, and conceptual clarity.

The findings suggest that reflective use of AI can help instructors anticipate conceptual difficulties, scaffold their sessions more effectively, and create learning environments that benefit a broad range of learners.
Positioning large language models as instructor-facing design companions, rather than as student-facing tutors, offers a low-risk, transferable model that could be applied to other theory-intensive STEM domains, including algorithms, discrete mathematics, and proof-based physics.

Future work could explore adapting this approach to multiple courses and institutions, comparing its impact across contexts, and experimenting with different prompting strategies to refine pedagogical insights.
Over time, the practice of AI-assisted instructional design could become as routine and valued as peer review of lecture materials, scaling high-quality teaching practices in challenging, conceptually dense subjects.

\vspace{10mm}

\bibliography{main}

\end{document}